# The WR population predicted by massive single star and by massive binary evolution.

**D. Vanbeveren** [1, 2] **, J. Van Bever** [3] **, H. Belkus** [1]

(1): Astrophysical Institute, Vrije Universiteit Brussel, Pleinlaan 2, 1050 Brussels, Belgium, dvbevere@vub.ac.be , hbelkus@vub.ac.be

(2): Mathematics department, Groep T, Leuven University Association, Vesaliusstraat 13, 3000 Leuven, Belgium

(3): Institute for Computational Astrophysics, Saint Mary's University, Halifax, NS, B3H 3C3, Canada, vanbever@ap.smu.ca

**Abstract**

We discuss differences between massive single star and massive close binary population number synthesis predictions of WR stars. We show that the WC/WN number ratio as function of metallicity depends significantly on whether or not binaries are included. Furthermore, the observed WC(+OB)/WN(+OB) number ratio in the Solar neighborhood seems to indicate that the WR mass loss rates are lower by another factor two compared to recently proposed clumping corrected formalisms. We then demonstrate that the observed lower luminosity distribution of single WN stars can be explained in a satisfactory way by massive single star evolutionary computations where the red supergiant phase is calculated using a stellar wind mass loss rate formalism that is based on recent observations.

## 1. Introduction

Comparing the results of theoretical population number synthesis (PNS) of massive stars with observed populations may help to determine values of uncertain parameters in the physics of processes governing massive star evolution such as interior mixing processes, stellar wind mass loss processes, the processes related to Roche lobe overflow (RLOF) in binaries etc. The first numerical PNS codes were developed in the 1980s and mainly used to study binary populations (e.g. Kornilov and Lipunov, 1983; Lipunov and Postnov, 1987; Dewey and Cordes, 1987; Eggleton et al., 1989; Tutukov and Yungel'son, 1987). Meanwhile PNS has become very popular and many studies have been published on various stellar objects, which are the product of binary star evolution.

The Brussels PNS code for massive single stars and binaries was described by Vanbeveren et al. (1998) and it was applied to investigate the Wolf-Rayet (WR) and O-type star populations as function of metallicity. This study accounted for evolutionary calculations where the stellar wind mass loss formalism used during core helium burning (CHeB) was based on empirical clumping corrected rates, but we did not account for a possible dependence of the WR mass loss rates on metallicity Z. One of the main conclusions was that the overall WR population in regions of continuous star formation depends critically on massive binary evolution. Van Bever and Vanbeveren (2003) presented evolutionary computations of massive single stars and binaries applying a metallicity dependent WR type stellar wind mass loss formalism. These results were used in order to investigate the effects of binaries on the evolution of WR type spectral features in starbursts. Also here it was concluded that binaries play an essential role in the interpretation of the observational data.

Of course, when observations allow the identification of WR stars which are indeed formed through the massive single star evolutionary scenario (notice that this is not trivial, Vanbeveren et al., 1998a), a comparison with theoretically predicted single WR star populations gives very valuable information as far as the evolution of massive single stars is concerned. A most recent example where this is done is the interesting study of Eldridge and Vink (2006) where massive single star evolutionary calculations are presented with a metallicity dependent stellar wind mass loss rate formalism during CHeB and where the observed and theoretically predicted WC/WN number ratios are compared. However, the observed numbers used by Eldridge and Vink do not separate binaries from single stars whereas the predicted numbers only account for single star evolution. The authors argue that binaries will only marginally alter these predicted numbers, but by reading their argumentation a word of caution seems appropriate The WR population in general, the WN/WC ratio in particular is critically affected by the stellar wind mass loss during CHeB and we agree with the authors that the effect of binarity on stellar wind mass loss during CHeB may be similar to the effect of rotation. However, the main effect of binarity on the WR population is NOT related to the latter but is related to the mass loss process in the pre-WR phase. Single stars become hydrogen deficient CHeB stars due to pre-WR stellar wind mass loss (during core hydrogen burning, during the luminous blue variable (= LBV) phase and during the red supergiant (= RSG) phase) and the moment during CHeB where the star becomes a WR star depends on these mass loss rates. However, due to the combined action of pre-RLOF stellar wind mass loss and

Roche lobe overflow mass loss, a massive primary becomes a hydrogen deficient CHeB star ALWAYS at the BEGINNING of its CHeB phase. As far as WR population synthesis is concerned, the latter process makes the difference between single stars and binary components.

Hamann et al. (2006) investigated 63 Galactic (mostly single) WN stars (a mixture of WNL and WNE types, e.g.. WN with and without hydrogen) with the Potsdam WR atmosphere grid models that account for wind homogeneities (clumping, they use a clumping factor D = 4) and line blanketing by iron and other iron-group elements. The resulting stellar parameters (listed for all objects) are then compared to population predictions using massive single star evolutionary models calculated with and without the effects of stellar rotation by the Geneva group (Meynet and Maeder, 2003). The authors conclude that independent from whether or not rotation is included, the correspondence is very poor; especially the lower luminosity hydrogen poor WNE population is not well reproduced.

The following two topics are the scopes of the present paper:

A. the effects of binaries on WR statistics are still underestimated in many studies and therefore we further highlight the differences between WR single star and binary population synthesis
B. we discuss uncertainties in processes which critically affect the evolution of massive stars and which critically determine theoretical WR star population synthesis, especially of the lower luminosity WR stars.

## 2. The formation and evolution of massive hydrogen deficient CHeB stars

Once a massive hydrogen deficient CHeB star is formed, its further evolution will be governed by WR-like stellar wind mass loss. WR mass loss rate formalisms, which account for the effects of clumping were presented by Vanbeveren et al. (1998a) and by Nugis and Lamers (2000). In table 1 we illustrate that both relations yield very similar evolutionary results. To illustrate the effects of a metallicty (Z) dependence we adopt a $\sqrt{Z}$ relation during the whole CHeB. Vink and de Koter (2005) discussed the Z-dependence of the WR mass loss rate and they propose a more complex relation that the one used here. This relation should obviously apply for WR single stars and for WR binary components. Since it is one of the scopes of the present paper in order to illustrate differences between single star and binary WR population synthesis, the simple $\sqrt{Z}$ relation is more than sufficient and this will be illustrated by comparing results calculated with a more complex Z-dependence promoted by Vink and de Koter.

## 2.a. Massive close binaries

Let us first recall a definition for a WR star that is frequently used in population studies:

*'A massive star will be classified as a WR star when $X_{surface}$ < 0.4 and log $T_{eff}$ > 4.0'.*

It can be considered as a general theorem that

> when a massive star (with initial ZAMS mass ≥ 10 $M_o$) is the primary (= the mass loser) of an interacting binary (binary period ≤ 10 years)
> and
> when during the evolution of the binary both components do not merge due to stellar wind mass loss followed by the Roche lobe overflow/common envelope process the massive primary will become a hydrogen deficient ($X_{surface}$ ≤ 0.3) star already at the beginning of its CHeB phase (e.g., Vanbeveren et al, 1998b, c, and references therein). The latter is true independent from the initial mass ratio or period of the binary.

When this theorem is combined with the WR-definition given above without further restriction, we would be forced to conclude that a primary of an interacting binary with an initial mass as low as 10 $M_o$ will become a WR star and this sounds rather unrealistic. The WR definition therefore needs some refinement and the following is a logical EXTRA criterion, i.e. a hydrogen deficient CHeB binary component will be classified as a WR star when its luminosity is larger than some minimum value. Because of the suspected tight mass-luminosity relation of hydrogen deficient massive CHeB stars (Vanbeveren and Packet, 1979; Langer, 1989) the latter can be rephrased using the mass instead of the luminosity. The orbital solutions of the known WR+OB binaries in combination with realistic OB-type star mass-luminosity calibrations, reveal that the minimum WR mass ($M_{min}$) is 8 $M_o$ (Vanbeveren et al., 1998a, b, c). The results of the present paper are calculated using the latter extra WR condition except when an alternative one is explicitely mentioned.

Let us summarize the advantage of studying the WR population in WR+OB binaries. Accounting for the general theorem given above it follows that the theoretically predicted population of WR stars in WR+OB binaries is independent from the initial mass ratio and period distribution of massive binaries, and of course it is largely independent from uncertainties of pre-WR mass loss rate formalisms like the one of RSGs. This is in particular true for the number ratio WC(+OB)/WN(+OB). The latter ratio then mainly depends on the mass loss rate during CHeB and can be considered as one of the best WR mass loss rate formalism indicators (remind that this is not true for the single WR star ratio since the latter depends on the RSG and LBV mass loss rate of massive stars which are still quite uncertain). It also means that to study the theoretically predicted population of WR stars in WR+OB binaries, it is sufficient to generate a population of hydrogen deficient ($X_{atm} < 0.3$) zero age CHeB stars and to calculate their further evolution as if they were single stars.

Table 1 lists the evolutionary parameters of (Galactic) hydrogen deficient ($X_{surface} \leq 0.3$) CHeB binary components for different mass loss rate prescriptions. Using a Salpeter type IMF for primaries of interacting close binaries, we also give the theoretically expected (Galactic) binary ratio WC(+OB)/WN(+OB). The observed sample of Galactic WR+OB binaries yields a ratio $\leq 0.6$ which is a factor two smaller that the expected value. Of course we are aware of the limitations due to small number statistics, however table 1 also gives the results when the CHeB mass loss rate is decreased by a factor 2 compared to the clumping corrected formalism of Nugis and Lamers (2000) or the one proposed by Vanbeveren et al. (1998a) and it is interesting to notice that the corresponding WC(+OB)/WN(+OB) ratio now matches the observed ratio. We consider this as evidence that the stellar wind mass loss rates of Galactic WR stars may be smaller by a factor two than predicted by these two formalisms.

We computed the evolution of massive CHeB binary components for Z = 0.004 and for Z = 0.001 and to illustrate the effect of the value of $M_{min}$ we repeated our calculations but with $M_{min}$ = 5 Mo. In figure 1a we plot the WC(+OB)/WN(+OB) number ratio as function of Z for both values of $M_{min}$, and we compare with the single star prediction (scheme D of Eldridge and Vink, 2006, which corresponds to the CHeB mass loss scheme used here). As can be noticed the binary results differ significantly from the single star prediction (the difference is a factor 1.5 at Solar and supersolar Z, up to a factor 2 for subsolar Z), contradicting the argumentation of

Eldridge and Vink outlined in the introduction of the present paper. A closer inspection reveals that the difference is mainly due to the differences between massive single star and massive binary evolution in the mass range 25 Mo and 40 Mo. One could be inclined to impute this difference to the chosen WR mass loss-Z-dependence. However, it is clear that once a Z-relation is chosen, it applies for WR single stars and for WR binary components. It can therefore be expected that the relative difference between single and binary WR stars only slightly depends on this Z-relation. This is illustrated in figure 1b showing results which correspond to scheme B of Eldridge and Vink. Figure 1a finally illustrates that as far as the WR binary population synthesis is concerned, a most important parameter is the adopted minimum mass $M_{min}$. Accurate WR binary orbit observations will further help to constrain this parameter and improve population models.

**2.b. Massive single stars**

The formation of single massive hydrogen deficient CHeB stars depends critically on the stellar wind mass rates during previous phases. For stars with initial mass ≥ 40 Mo, LBV type mass loss rates are decisive, however for lower masses the RSG mass loss rates become increasingly more important. The RSG evolutionary phase of massive single stars is generally computed by using the de Jager et al. (1988) formalism (e.g., Meynet and Maeder, 2003; Eldridge and Vink, 2006). However, an update may be necessary. In figure 2 we compare RSG rates predicted with the de Jager et al. (1988) formalism and recent observations of Van Loon et al. (2005). It is clear that the differences are substantial. On this figure we also plot earlier RSG mass loss rate determinations for LMC supergiants of Reid et al. (1990) Also these rates point towards an alternative RSG formalism. Vanbeveren (1995) proposed such an alternative and the effect on massive single star evolution was studied by Vanbeveren et al. (1998a, b) and Salasnich et al., (1999). One of the main consequences of these higher rates is that single stars with an initial mass between 20 Mo and 40 Mo lose most of their hydrogen rich layers during their RSG phase and become WR stars. Using these computations we can investigate whether or not the recent observations of Hamann et al. (2006) of the lower luminosity WR stars are better explained. We performed the following exercise. From the list of Hamann et al. we selected the single WN stars with a luminosity Log L/Lo ≤ 6 and we count the number of WNE and WNL stars in the log $L/L_o$ intervals [5.2-5.4], [5.4-5.6], [5.6-5.8] and [5.8-6.0]. The resulting histogram is given in Figure 3. Using a Salpeter IMF and accounting for the WR evolutionary timescales of single star evolutionary calculations, it is straightforward in order to predict the number of single WR stars in these luminosity bins. Figure 3 also shows the theoretically predicted histogram when the Geneva massive single star tracks are used with and without rotation (note that these tracks are calculated with the standard de Jager et al. RSG mass loss rate formalism) and when our evolutionary tracks are used with the new RSG mass loss formalism. We confirm the conclusion of Hamann et al. that massive single star evolution calculated with the standard RSG mass loss rate formalism does not reproduce the observations. However, the predictions with tracks with the new RSG rates are significantly closer to the observations. We therefore strongly advice that in existing single star evolutionary codes the old RSG mass loss rate formalism is replaced by a formalism that accounts for recent observations. Since RSG mass loss is very important for the evolution of single stars with an initial mass lower than 40 Mo, extensive RSG mass loss studies may prove to be very valuable.

3. **Conclusions**

In the present paper we discussed the effect of binaries and the effect of uncertainties in the red supergiant stellar wind mass loss formalism on the theoretically predicted distribution of WR stars. We conclude

a. the WC/WN number ratio as function of metallicity predicted by massive binary evolution may be significantly different from the one predicted by massive single star evolution
b. the WC(+OB)/WN(+OB) number ratio in the Solar neighborhood may indicate that the stellar wind mass loss rates of WR stars may be smaller by another factor 2 compared to recent clumping corrected rates
c. the red supergiant stellar wind mass loss rates of stars with an initial mass < 40 Mo may be much larger than predicted by the formalism that is used in most stellar evolutionary codes;
d. The distribution of the number of single WR stars as function of luminosity is much better reproduced by a population model that uses evolutionary computations that account for the higher RSG mass loss rates.

**Table and figure captions**

|          | Mbegin | Mend  | T_WR | T_WNL | T_WNE | T_WC | Mzams  | WC/WN |
|----------|--------|-------|------|-------|-------|------|--------|-------|
| **Mdot VB** | 60.00  | 19.20 | 3.51 | 0.04  | 0.38  | 3.09 | 103.91 | 1.36  |
|          | 50.00  | 16.30 | 3.75 | 0.05  | 0.44  | 3.26 | 88.28  |       |
|          | 40.00  | 13.50 | 4.09 | 0.07  | 0.52  | 3.50 | 72.66  |       |
|          | 30.00  | 10.60 | 4.62 | 0.11  | 0.72  | 3.79 | 57.03  |       |
|          | 20.00  | 7.60  | 5.24 | 0.21  | 1.31  | 3.72 | 41.41  |       |
|          | 10.00  | 4.40  | 1.90 | 0.77  | 1.13  | 0.00 | 25.78  |       |
|          | 8.00   | 3.70  | 0.00 | 0.00  | 0.00  | 0.00 | 22.66  |       |
| **Mdot NL** | 60.00  | 22.49 | 3.69 | 0.04  | 0.23  | 3.42 | 103.91 | 1.37  |
|          | 50.00  | 20.12 | 3.89 | 0.05  | 0.27  | 3.57 | 88.28  |       |
|          | 40.00  | 17.54 | 4.16 | 0.07  | 0.35  | 3.74 | 72.66  |       |
|          | 30.00  | 14.65 | 4.57 | 0.11  | 0.53  | 3.93 | 57.03  |       |
|          | 20.00  | 11.23 | 5.34 | 0.23  | 1.13  | 3.98 | 41.41  |       |
|          | 10.00  | 6.87  | 2.33 | 0.96  | 1.37  | 0.00 | 25.78  |       |
|          | 8.00   | 5.80  | 0.00 | 0.00  | 0.00  | 0.00 | 22.66  |       |
| **Mdot/2** | 60.00  | 33.40 | 3.17 | 0.08  | 0.75  | 2.34 | 103.91 | 0.44  |
|          | 50.00  | 27.98 | 3.38 | 0.10  | 0.85  | 2.43 | 88.28  |       |
|          | 40.00  | 22.62 | 3.67 | 0.14  | 1.01  | 2.52 | 72.66  |       |
|          | 30.00  | 17.30 | 4.11 | 0.21  | 1.39  | 2.51 | 57.03  |       |
|          | 20.00  | 11.94 | 4.92 | 0.41  | 2.44  | 2.07 | 41.41  |       |
|          | 10.00  | 6.40  | 3.54 | 1.47  | 2.07  | 0.00 | 25.78  |       |
|          | 8.00   | 5.26  | 0.00 | 0.00  | 0.00  | 0.00 | 22.66  |       |
| **Mdot/4** | 60.00  | 44.58 | 3.04 | 0.15  | 1.46  | 1.43 | 103.91 | 0.05  |
|          | 50.00  | 37.22 | 3.22 | 0.19  | 1.64  | 1.39 | 88.28  |       |
|          | 40.00  | 29.87 | 3.49 | 0.27  | 1.92  | 1.30 | 72.66  |       |
|          | 30.00  | 22.59 | 3.89 | 0.41  | 2.59  | 0.89 | 57.03  |       |
|          | 20.00  | 15.29 | 4.62 | 0.79  | 3.83  | 0.00 | 41.41  |       |
|          | 10.00  | 7.89  | 6.26 | 2.72  | 3.54  | 0.00 | 25.78  |       |
|          | 8.00   | 6.40  | 0.00 | 0.00  | 0.00  | 0.00 | 22.66  |       |

**Table 1:** The CHeB evolution of massive RLOF remnants with post-RLOF mass M (in Mo) for different WR mass loss rate formalisms (Mdot VB = the formalism proposed by Vanbeveren et al., 1998; Mdot NL = the formalism proposed by Nugis and Lamers, 2000; Mdot/2 and Mdot/4 correspond to the case where the NL mass loss rate is reduced by a factor 2 and 4). We give the mass at the end of CHeB (Mend in Mo), the WR lifetime (T_WR), the WNL, WNE and WC lifetime (T_WNL, T_WNE and T_WC) (all lifetimes are in $10^5$ yr.), the corresponding ZAMS mass (Mzams in Mo) and the resulting WC/WN binary number ratio using a Salpeter IMF.

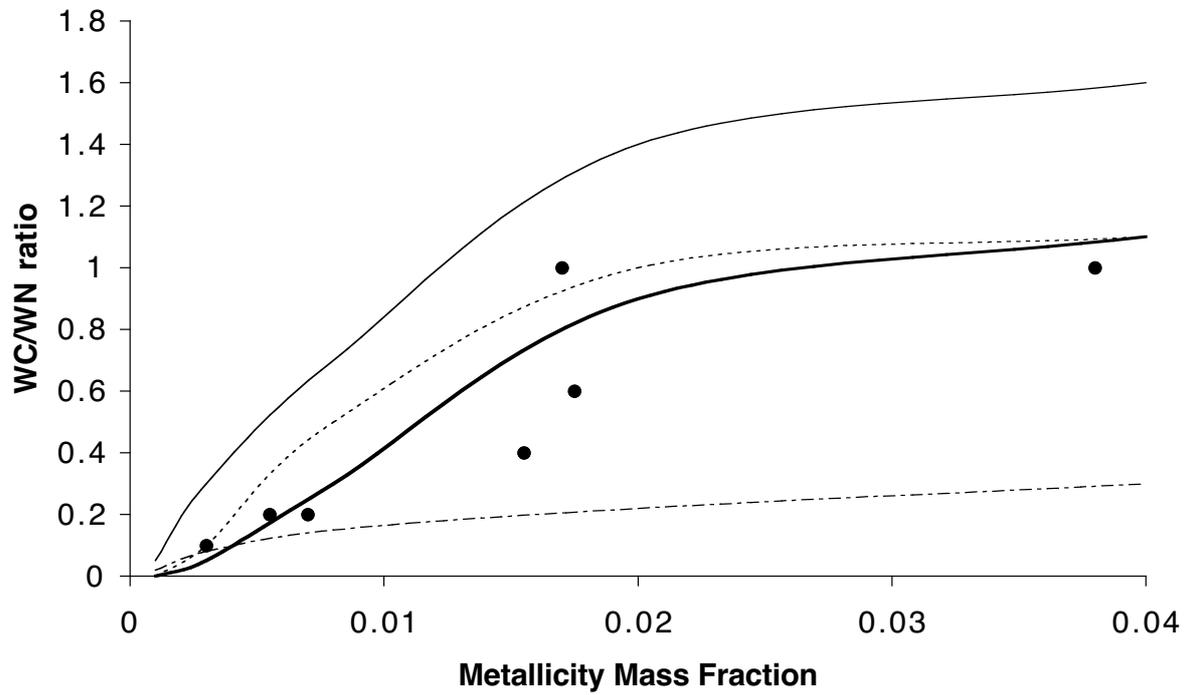

**Figure 1a:** The WC/WN number ratio as function of metallicity mass fraction Z. The dots are observed values (see Eldridge and Vink, 2006), the thin line gives the ratio for primaries of binaries when the minimum mass of WR stars Mmin = 8 Mo, the dashed-dotted line gives the ratio for primaries of binaries when Mmin = 5 Mo, the thick line is similar to the thin line but the WR mass loss is reduced by a factor 2, the dashed line gives the predicted WC/WN ratio for single stars of Eldridge and Vink (2006).

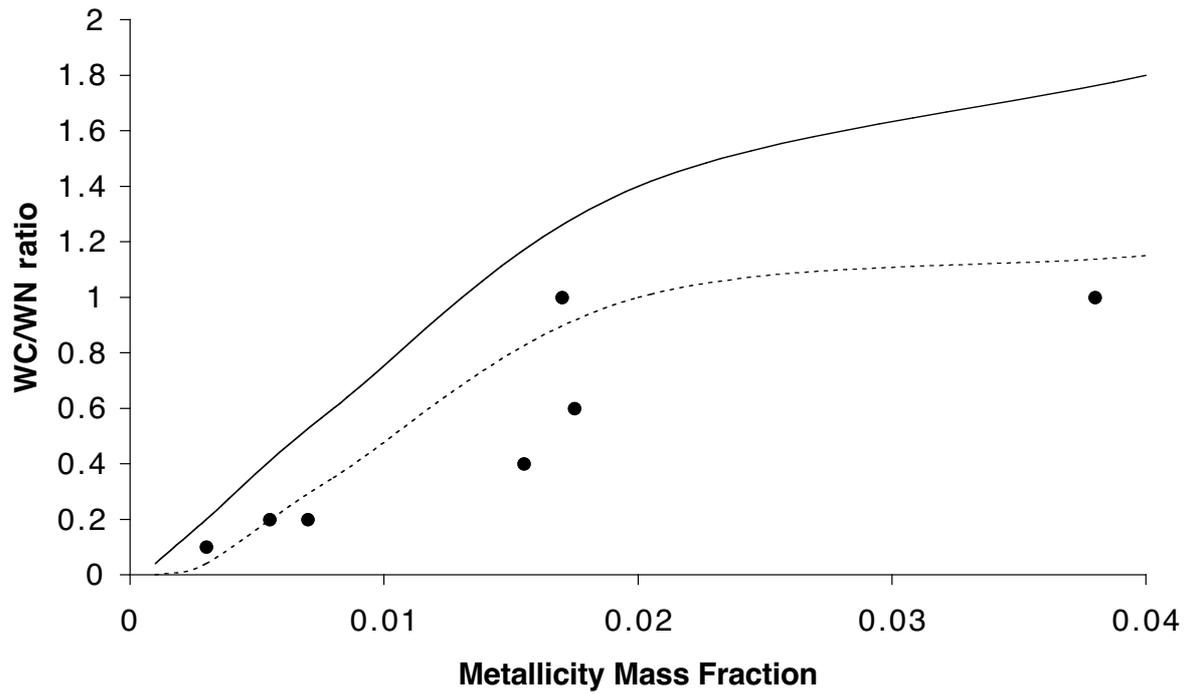

**Figure 1b:** The lines have the same meaning as in figure 1a but we use evolutionary computations during CHeB where the WR-Z mass loss rate relation is used corresponding to scheme B of Eldridge and Vink (2006).

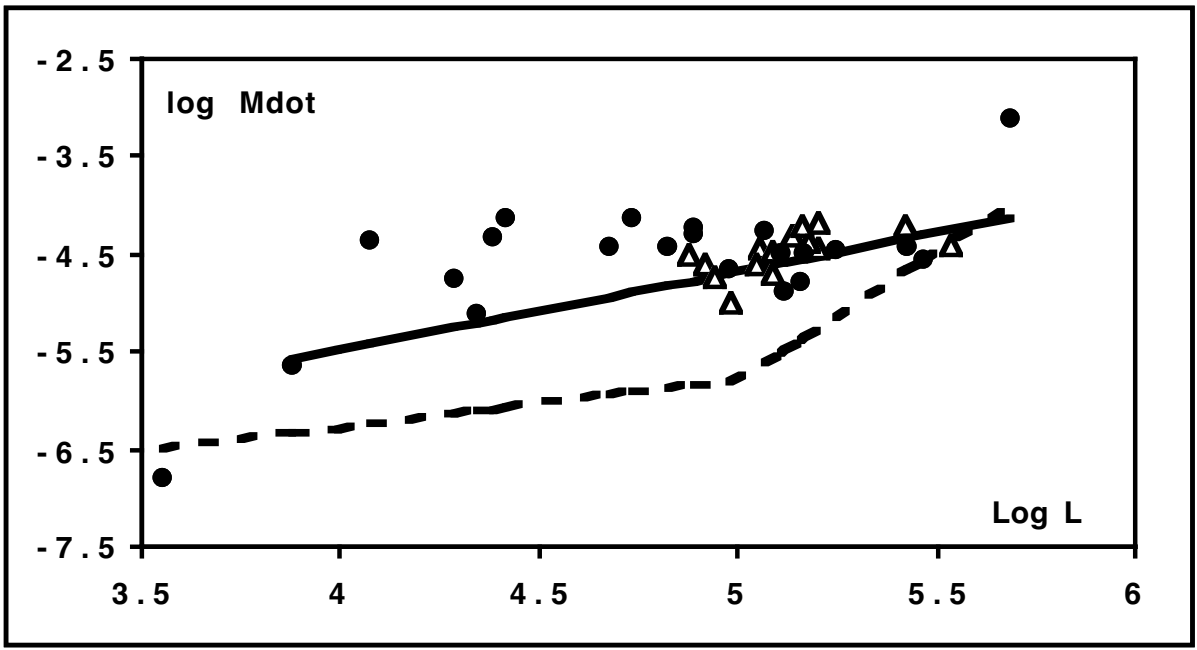

**Figure 2:** The RSG mass loss formalism of de Jager et al. (1988) (dashed line) and the one proposed by Vanbeveren et al. (1998) (full line) compared to observations of van Loon et al. (2005) (black dots) and of Reid et al. (1990) (open triangles).

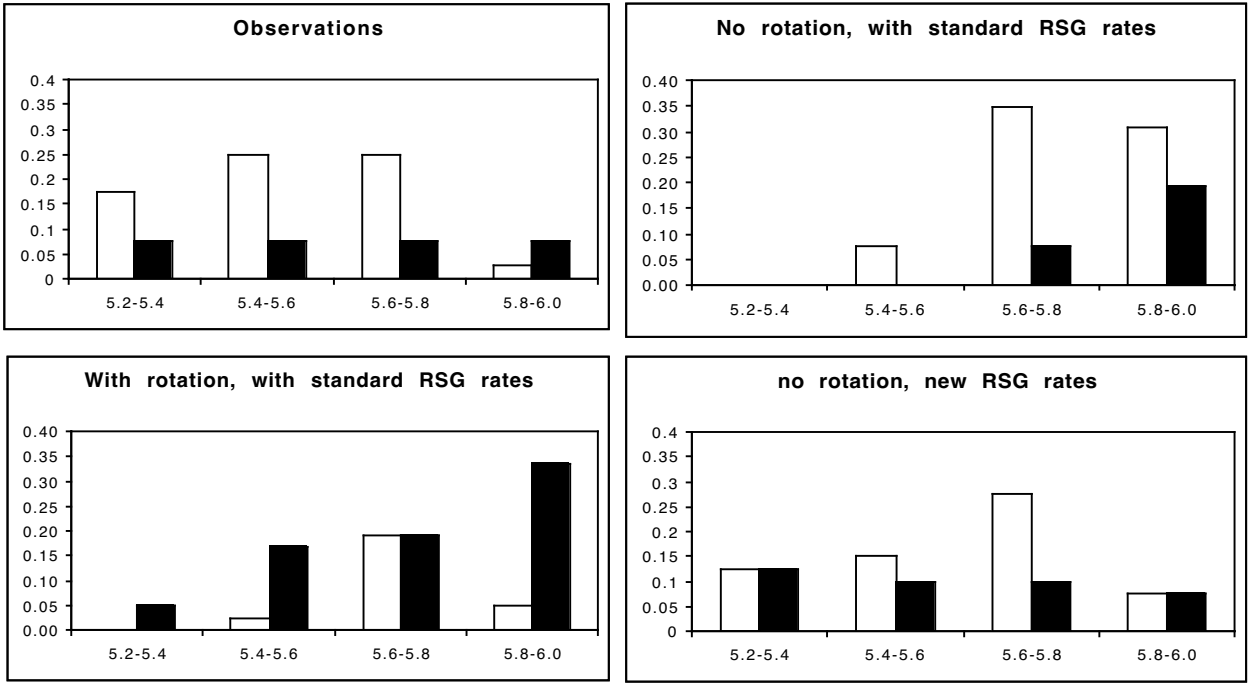

**Figure 3:** The number distribution as function of luminosity of WNE stars (white histogram) and of WNL stars (black histogram). We compare the observations of Hamann et al. (2006) with predictions using the Geneva tracks (with and without rotation, standard = de Jager et al. mass loss rates during the RSG phase) and with predictions using tracks with the alternative RSG mass loss rate formalism.